\documentclass{aa}  

\usepackage{graphicx,epsfig}
\usepackage{footnote}
\usepackage{txfonts}
\usepackage{color}
\definecolor{redHL}{rgb}{1.0,0.5,0.5}

\begin{document} 

\title{Unveiling the enigma of ATLAS17aeu\thanks{Based on observations made with Copernico, the TNG (under programme A34TAC$\_$24), the GTC (under programmes GTCMULTIPLE2D-16B and GTCMULTIPLE2G-17A), the LBT (under programme 2016$\_$2017$\_$19), and with the {\it HST} (under programme GO14270) telescopes.}}

   %\subtitle{}

%\author{GRAWITA collaboration et al....}
\author{A. Melandri\inst{1}, A. Rossi\inst{2}, S. Benetti\inst{3}, V. D'Elia\inst{4,5}, S. Piranomonte\inst{5}, E. Palazzi\inst{2}, A. J. Levan\inst{6}, M. Branchesi\inst{7,8}, A. J. Castro-Tirado\inst{9}, P. D'Avanzo\inst{1}, Y.-D. Hu\inst{9}, G. Raimondo\inst{10}, N. R. Tanvir\inst{11}, L. Tomasella\inst{3}, L. Amati\inst{2}, S. Campana\inst{1}, R. Carini\inst{3}, S. Covino\inst{1}, F. Cusano\inst{2}, M. Dadina\inst{2}, M. Della Valle\inst{12,9}, X. Fan\inst{13}, P. Garnavich\inst{14}, A. Grado\inst{12}, G. Greco\inst{15}, J. Hjorth\inst{16}, J. D. Lyman\inst{6}, N. Masetti\inst{2,17}, P. O'Brien\inst{11}, E. Pian\inst{2}, A. Perego\inst{18,19},  R. Salvaterra\inst{20}, L. Stella\inst{5}, G. Stratta\inst{15}, S. Yang\inst{3}, A. di Paola\inst{5}, M. D. Caballero-Garc\'ia\inst{21}, A. S. Fruchter\inst{22}, A. Giunta\inst{5}, F. Longo\inst{23}, M. Pinamonti\inst{24}, V. V. Sokolov\inst{25}, V. Testa\inst{5}, A. F. Valeev\inst{25}, and E. Brocato\inst{5} on behalf of the Gravitational Wave InAf TeAm (GRAWITA)}

\institute{INAF - Osservatorio Astronomico di Brera, Via E. Bianchi 46, I-23807, Merate (LC), Italy\\
		\email{andrea.melandri@brera.inaf.it}
		\and INAF - Osservatorio di Astrofisica e Scienza dello Spazio di Bologna, Via Piero Gobetti 93/3, I-40129, Bologna, Italy 
        \and INAF - Osservatorio Astronomico di Padova, Vicolo dell'Osservatorio 5, I-35122, Padova, Italy
		\and ASI - Science Data Centre, Via del Politecnico snc, I-00133, Roma, Italy
		\and INAF - Osservatorio Astronomico di Roma, Via di Frascati, 33, I-00040 Monteporzio Catone, Italy
		\and Department of Physics, University of Warwick, Coventry, CV4 7AL, UK
		\and Gran Sasso Science Institute, Viale F. Crispi 7, I-67100, L'Aquila, Italy
		\and INFN - Laboratori Nazionali del Gran Sasso, I-67100 L'Aquila, Italy
		\and Instituto de Astrof\'{i}sica de Andaluc\'{i}a (IAA-CSIC), Glorieta de la Astronomıa, s/n, E-18008, Granada, Spain
        \and INAF - Osservatorio Astronomico d'Abruzzo, Via Mentore Maggini, I-64100, Teramo, Italy
		\and Department of Physics and Astronomy, University of Leicester, Leicester, LE1 7RH, UK
        \and INAF - Osservatorio Astronomico di Capodimonte, Salita Moiariello 16, I-80131, Napoli, Italy 
		\and Steward Observatory, The University of Arizona, 933 North Cherry Avenue, Tucson, AZ 85721-0065, USA
        \and Department of Physics, University of Notre Dame, Notre Dame, IN 46556, USA
       \and Urbino University, Via Santa Chiara 27, I-61027, Urbino, Italy
        \and Dark Cosmology Centre, Niels Bohr Institute, University of Copenhagen, Juliane Maries Vej 30, DK-2100, Copenhagen, Denmark 
		\and Departamento de Ciencias F\'{i}sicas, Universidad Andr\'{e}s Bello, Fern\'{a}ndez Concha 700, Las Condes, Santiago, Chile
 		\and INFN - Sezione Milano Bicocca, Gruppo Collegato di Parma, Parco Area delle Scienze 7/A, I-43124, Parma, Italy
        \and Dipartimento di Fisica, Universit\'{a} degli Studi di Milano Bicocca, Piazza della Scienza 3, I-20126, Milano, Italy
        \and INAF - IASF Milano, Via E. Bassini 15, I-20133, Milano, Italy
		\and Astronomical Institute, Academy of Sciences of the Czech Republic, Bocni II 1401, CZ-141 00 Prague, Czech Republic
        \and Space Telescope Science Institute, 3700 San Martin Drive, Baltimore, MD 21218, USA
        \and Department of Physics, University of Trieste, and INFN, sezione di Trieste, Via Valerio 2, I-34127, Trieste, Italy
        \and INAF - Osservatorio Astronomico di Torino, Strada Osservatorio 20, I-10025 Pino Torinese, Italy 
        \and Special Astrophysical Observatory, Nizhnij Arkhyz, Karachai-Cherkessian Republic, 369167, Russia
}

   \date{}

    \titlerunning{The enigma of ATLAS17aeu}
   \authorrunning{Melandri et al.}

%
% \abstract{}{}{}{}{} 
% 5 {} token are mandatory
 
  \abstract
   {}
   {The unusual transient ATLAS17aeu was serendipitously detected within the sky localisation of the gravitational wave trigger GW\,170104. The importance of a possible association with gravitational waves coming from a binary black hole merger led to an extensive follow-up campaign, with the aim of assessing a possible connection with GW\,170104.}
   {With several telescopes, we carried out both photometric and spectroscopic observations of ATLAS17aeu, for several epochs, between $\sim 3$ and $\sim 230$ days after the first detection.}
   {We studied in detail the temporal and spectroscopic properties of ATLAS17aeu and its host galaxy. We detected spectral features similar to those of a broad lined supernova superposed to an otherwise typical long-GRB afterglow. Based on analysis of the optical light curve,  spectrum and host galaxy SED, we conclude that the redshift of the source is probably $z \simeq 0.5 \pm 0.2$.}
   {While the redshift range we have determined is marginally compatible with that of the gravitational wave event, the presence of a supernova component and the consistency of this transient with the E$_{\rm p}$-E$_{\rm iso}$ correlation support the conclusion that ATLAS17aeu was associated with the long gamma-ray burst GRB\,170105A. This rules out the association of the GRB\,170105A/ATLAS17aeu transient with the gravitational wave event GW\,170104, which was due to a binary black hole merger.}

   
   \maketitle

%-------------------------------------------------------------------

\section{Introduction}
The first direct observation of gravitational waves by the Advanced LIGO interferometers came from the coalescence of a binary system of black holes \citep{LVC2016a}, opening the era of gravitational-wave astronomy. Gravitational-wave signals from binary systems of black holes were detected several times during the first and second run of observations of the advanced detectors \citep{LVC2017a,LVC2017b,LVC2017c,LVC2016b}. They provided us with information about their rate and mass distribution, and probed their formation and evolution. The first gravitational-wave signal from the inspiral of a binary neutron star system GW\,170817 \citep{LVC2017d} was observed on August 17, 2017 by the Advanced LIGO and Virgo network, starting the era of multi-messenger astronomy \citep{Abbott2017f}.

On January 4, 2017 at 10:11:58.6 UTC the Advanced LIGO detectors revealed the signal from a binary black-hole coalescence, GW\,170104 \citep{LVC2017a}. The system was made of component black-hole of masses ${31.2}_{-6.0}^{+8.4}$ M$_{\rm \odot}$ and ${19.4}_{-5.9}^{+5.3}$ M$_{\rm \odot}$ (at the 90\% confidence level) at a luminosity distance of $880_{-390}^{+450}$ Mpc corresponding to a redshift of $z={0.18}_{-0.07}^{+0.08}$ \citep{LVC2017a}. An alert with an initial source localization ($\sim$ 1600 deg$^{2}$ at the 90\% confidence level) was distributed to collaborating astronomers \citep{gw170104}. During the electromagnetic counterpart follow-up search, the ATLAS and Pan-STARRS surveys discovered ATLAS17aeu \citep{tonry}, 23.1 hr after GW\,170104, which was a rapidly fading transient within the inner 16\% sky-localization probability contour (see Fig. {\ref{FigLoc}). The transient, with a decay similar to a GRB afterglow, was also detected in X-rays by {\it Swift} and in the radio at 6 and 15 GHz by the VLA and the AMI large array, respectively \cite{evans17a,evans17b,corsi17,mooley17}. 

By fitting a power law to the optical decay, the time zero was found consistent with the gamma-ray burst GRB\,170105A \citep{Kasliwal2017} detected by the POLAR instrument onboard the Chinese space laboratory Tiangong-2 \citep{polar}, {\it AstroSat}-CZTI \citep{as1}, {\it Konus}-Wind, and {\it INTEGRAL}-SPIACS \citep{ipn} 20.04 hr after GW\,170104. Temporal and spatial consistency led to the conclusion that ATLAS17aeu was the afterglow of GRB\,170105A and unrelated to GW\,170104 \citep{bhalerao}. Considering all the multi-wavelength observations of ATLAS17aeu, \citet{stalder17} concluded that the GRB\,170105A is compatible with a classical long-GRB at redshift 1 $\lesssim z \lesssim$ 2.9 and that ATLAS17aeu is statistically likely the associated afterglow. However, they evaluated a small but non-negligible probability of association of ATLAS17aeu and the GW signal, which only a direct redshift measurement of the host galaxy of ATLAS17aeu could exclude.

%\citet{stalder17} evaluate a small but non negligible probability of association of the GRB and GW signal, but since the properties of the afterglow and host galaxy are compatible with a classical GRB at typical redshift (1 $\lesssim z \lesssim$ 2.9), they conclude that the two events are probably not associated and only a direct measurement of the host redshift could clarify this point.

We present optical observations of ATLAS17aeu transient and its possible host galaxy taken with the 1.8-m Asiago Copernico telescope, the 3.6-m Telescopio Nazionale Galileo (TNG), the 8.4-m Large Binocular telescope (LBT), the 10.4-m Gran Telescopio Canarias (GTC), and the {\em Hubble Space Telescope} ({\it HST}) over 230 days from the GRB\,170105A trigger time. Together with the radio and X-ray observations, the connection between the transient ATLAS17aeu and the long gamma-ray burst GRB\,170105A is discussed. 

Throughout the paper, distances are computed assuming a $\Lambda$ CDM-Universe with H$_{\rm 0}$ = 71 km s$^{-1}$ Mpc$^{-1}$, $\Omega_{\rm m}$ = 0.27, and $\Omega_{\rm \Lambda}$ = 0.73 \citep{larson,koma}. Magnitudes are in the AB system and errors are at a 1$\sigma$ confidence level.

\section{ATLAS17aeu}

The rapidly fading transient ATLAS17aeu \citep{tonry}, identified within the localization of GW\,170104 \citep{gw170104}, was only 20$^{\prime\prime}$ away from the SDSS galaxy J091312.36+610554.2, with a spectroscopic redshift ($z \sim 0.2$) consistent with the distance inferred for GW\,170104. Considering this galaxy as possible host of ATLAS17aeu, the position and distance consistency of ATLAS17aeu and GW\,70104 led to many multi-wavelength observations to probe the possible association of ATLAS17aeu with the gravitational signal. 

   \begin{figure}
   \centering
    \includegraphics[width=9.0cm,height=5.5cm]{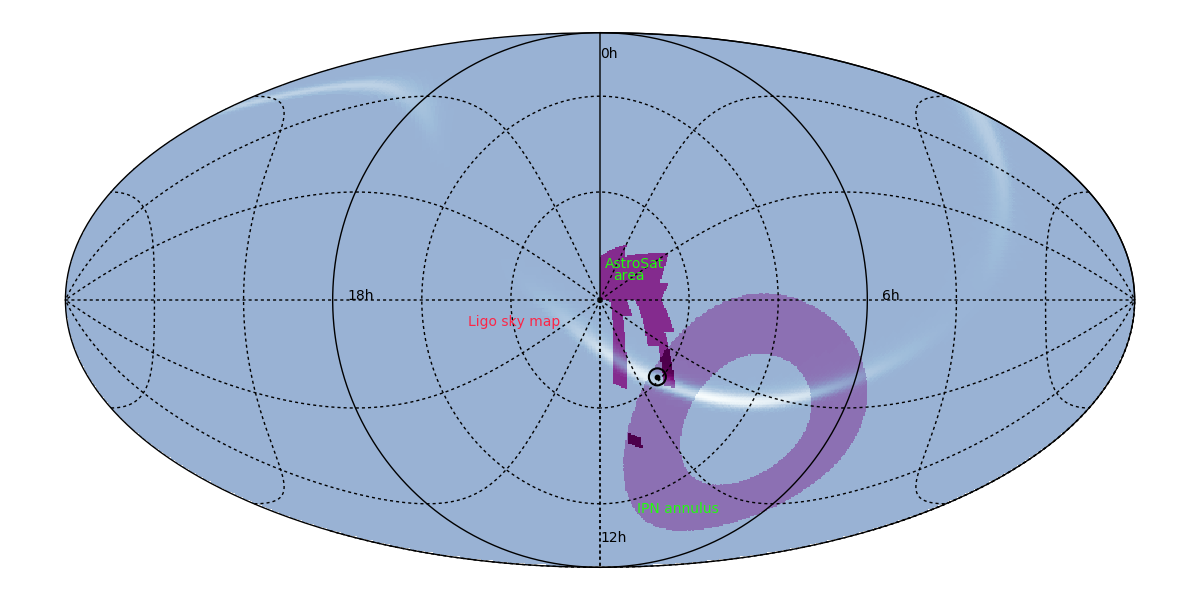}
   \caption{The mollweide projection for localisation area of ATLAS17aeu (black circle). We report the AstroSat CZTI localisation area (dark purple) at 1$\sigma$ confidence level and the IPN triangulation annulus (light purple) at 3$\sigma$ confidence level for GRB\,170105A. The final LVC sky map for GW\,170104 is shown in white.}
    \label{FigLoc}%
    \end{figure}

Within the GRAWITA\footnote{GRAvitational Wave Inaf TeAm: https://www.grawita.inaf.it/} framework, we monitored the light curve of the source between 1.65 and 88.7 days after the GRB trigger with the Asiago and TNG telescopes. In addition, we obtained two multi-filters epochs ($griz$) with the GTC telescope between 3.92 and 78.7 days after the GRB trigger. Finally, we observed the field with the LBT telescope in imaging mode ($gri$) at $\sim 104$~days after the gamma-ray burst trigger, GRB\,170105A. Image reduction was carried out following standard procedures and the optical data were calibrated using a common set of selected catalogued stars of the SDSS catalog present in the field of view. 

Two spectra were also acquired with the GTC telescope, one at $\sim 3$~days and a second one secured at $\sim 116$~days after the burst event. GTC spectroscopy was carried out using the OSIRIS camera in slit mode, with the R1000B ($R=1000$, spectral range $3630 - 7500$ \AA) and R2500I ($R=2500$, spectral range $7330 - 10000$ \AA) grisms. The slit width was set to $1^{\prime\prime}$. The data were optimally extracted \citep{horne86} and reduced following standard procedures using ESO MIDAS\footnote{http://www.eso.org/projects/esomidas/} and IRAF\footnote{http://iraf.noao.edu/} software.

In addition, three further optical spectra were collected on April 14, 2017, on January 25, 2018, and on March 19, 2018 with LBT, using the two Multi-Object Double Spectrograph \citep[MODS, ][]{Pogge2010}. All observations were obtained in the spectral range 3200$-$9500~\AA\, with a 1$^{\prime\prime}$ slit ($R \sim$ 2000). MODS uses two red- and blue-optimized channels with a spectral range of $3500-6500$ \AA~ and $5000-10000$ \AA, respectively. The first two epochs were taken with the one grating for each channel which has the advantage to avoid a gap at $\sim5650$~\AA\ between the two channels but doubles the observing time. Therefore, the last epochs were taken with the dual grating mode in which the light is separated by a dichroic into red- and blue-channels. A 2x2 binning was set in the second epoch, but it caused read-out artifacts and thus in the final epoch we adopted a 1x2 binning. The last observation was obtained under the best conditions (seeing $\sim 0.7\arcsec$, airmass $1.1-1.2$) for a total exposure time of $4800$~s. Data reduction was performed at the Italian LBT Spectroscopic Reduction Center\footnote{http://www.iasf-milano.inaf.it/Research/lbt\_rg.html} by means of scripts optimized for LBT data. Steps of the data reduction of each two-dimensional spectral image are the correction for dark and bias, bad-pixel mapping, flat-fielding, sky background subtraction, and extraction of one-dimensional spectrum by integrating the stellar trace along the spatial direction. Wavelength calibration was obtained from the spectra of arc lamps, while calibration was obtained using catalogued spectrophotometric standards.

The location of ATLAS17aeu was subsequently observed with the {\it HST}-WFC on August 22, 2017 ($\sim$ 229 days after the burst event). At this point, observations were obtained in the UVIS arm F390W, F606W and the IR arm with F140W\footnote{HST observations have been cross-calibrated with the $g$, $r$, and J bands, respectively.}. Observations were reduced by {\tt astrodrizzle} in the standard fashion. At the location of ATLAS17aeu, we clearly detect a source in both F606W and F140W, but there is no detection in F390W (Fig. \ref{FigHST}). The position of the ATLAS17aeu transient is RA = 09:13:13.89, Dec = +61:05:32.54 with an error of 0.06$\arcsec$ .

For our UVIS observations, we measure the AB magnitudes (or upper limits) within a 0.1\arcsec ~aperture and correct them with the published encircled energy curves\footnote{\url{http://www.stsci.edu/hst/wfc3/phot_zp_lbn}}. We determine that F390W$>28.1$ mag ($3\sigma$) and F606W=27.64 $\pm$ 0.21 mag. In the IR we use a 0.2\arcsec ~aperture due to the poorer PSF, and measure F140W=25.87 $\pm$ 0.14 mag. There is no sign of extension in the images, and the sources appear point-like. However, at this faint magnitude the detection of extension is challenging. We consider the source located S-E with respect to ATLAS17aeu as its host galaxy. The separation between the two objects is $\sim$ 1.8 $\arcsec$ (Fig. \ref{FigHST}).

The summary of our photometric and spectroscopic observations is reported in Tables \ref{LogTab} and \ref{LogTab2}, respectively. Data have not been corrected for Galactic extinction \citep[E$_{\rm B-V} = 0.028$~mag,][]{Schlafly2011}.

~\\

   \begin{figure}[!ht]
   \centering
    \includegraphics[width=9.0cm,height=7.5cm]{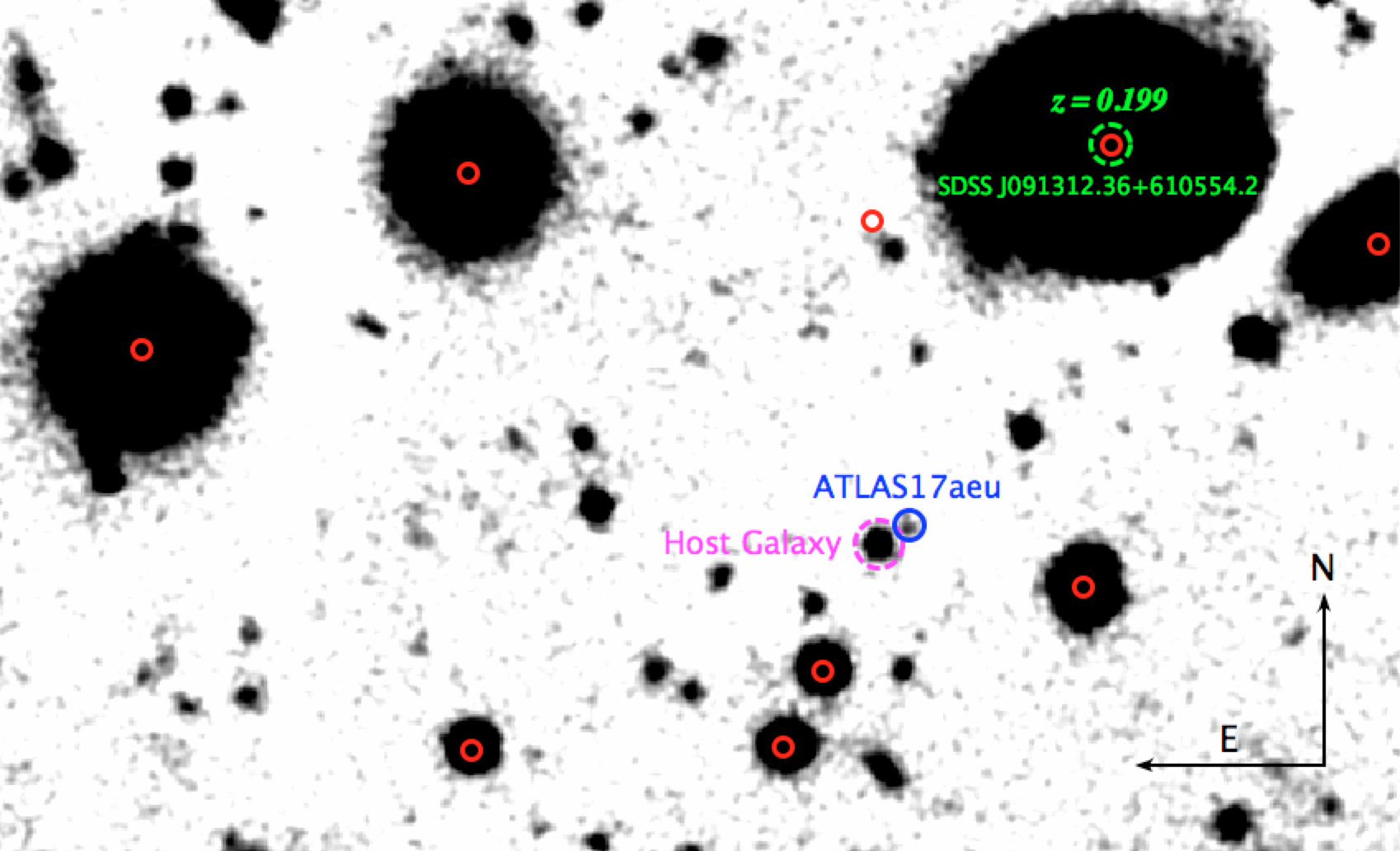}
   \caption{TNG image (field of view $\sim 1^{\prime} \times 1^{\prime}$) acquired at $\Delta t \sim$ 56 days, in the $r$ filter. The ATLAS17aeu position (blue circle) is $\sim 2^{\prime\prime}$ away from the centre of its host galaxy (magenta circle). The cataloged SDSS objects (red circles), including the J091312.36+610554.2 galaxy at $z \sim 0.2$ (green circle), are shown.}
    \label{FigCampo}%
    \end{figure}

\section{GRB\,170105A}

GRB\,170105A was detected at 06:14:07.0 UT (T$_{\rm 0}$, corresponding to MJD=57758.259803) with a total duration T$_{\rm 90}$ = 2.0 $\pm$ 0.5~s. The burst event was also detected by {\it INTEGRAL}-SPIACS, {\it Konus}-Wind, and {\it AstroSat}-CZTI \citep{as1} with a measured duration T$_{\rm 90} \sim$ 2.9~s. Its fluence derived from the {\it Konus}-Wind observation is S$_{[0.02-10~\rm MeV]} \sim 2.5 \times 10^{-6}$ erg cm$^{-2}$ and it displayed longer emission (with a duration of about 20 seconds) in the 18-70 keV soft channel of {\it Konus}-Wind \citep{ipn,stalder17}. 

In Fig. {\ref{FigLoc} we show the localisation areas ({\it AstroSat} and IPN) for this event, together with the LVC sky map for GW\,170104 and the most accurate position for ATLAS17aeu. As it can be seen ATLAS17aeu is slightly outside the 1$\sigma$ {\it AstroSat}/CZTI localisation area, well within the LVC probability contours. The temporal and spatial coincidence between ATLAS17aeu and GRB\,170105A indicated that the two events were most likely associated \citep{Kasliwal2017,as2,bhalerao} while it remained unclear the association with the GW\,170104 due to the lack of a firm ATLAS17aeu distance determination.

The GRB\,170105A fluence is consistent with a long-soft, under-energetic GRB. In fact, assuming the distance inferred for GW\,170104 ($z \sim 0.1$) the estimated isotropic energy of the GRB event would be E$_{\rm iso} \sim 5.8 \times 10^{49}$ erg, and at larger distances up to $z \approx$ 1, the isotropic energy remains still consistent with the faint end of the E$_{\rm iso}$  distribution for long GRBs \citep{nava,davanzo}.

   \begin{figure*}[!ht]
   \centering
    \includegraphics[width=18.0cm,height=6.0cm]{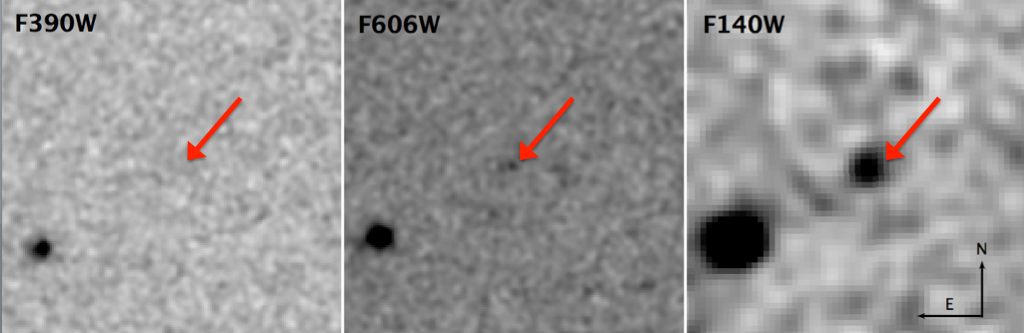}
   \caption{HST observations of the field of ATLAS17aeu (field of view = 5$\arcsec \times$ 5$\arcsec$). The red arrow indicates the location of the optical transient. The other object visible in that image is what we consider the host galaxy of ATLAS17aeu, reported also in Fig. \ref{FigCampo}. The offset between the two objects is $\sim$ 1.8 $\arcsec$.}
    \label{FigHST}%
    \end{figure*}

\section{Results and discussion}

\subsection{Temporal analysis}

The early time optical light curve of ATLAS17aeu (Fig. \ref{FigLC}) can be described by a single power-law decay ($\alpha_{\rm r} = 1.38\pm 0.02$). At a later time ($>$ 10 days) a significant deviation from that decay is detected, unveiling the presence of a possible supernova component (Fig. \ref{FigLC}). 

Many known under-energetic long-duration gamma-ray bursts and X-ray flashes have an associated highly stripped-envelope core-collapse supernova (Type Ib/c). At low redshifts ($z \lesssim 0.3$) the supernova component is well identified both photometrically and spectroscopically \citep{gala98,patat01,hj03,male04,ferrero06,pian06,cano11a,bufano12,mela12,schu14,mela14,delia15}, while at higher redshifts ($ 0.3 \lesssim z \lesssim 1$) the presence of the supernova is inferred from the detection of a re-brightening in the late afterglow light curve \citep{bloom99,ctg99,gala00,ct01,dv03,greiner03,zeh04,dv06,sode06,cano11b,sparre11,jin13}. The supernova origin for the re-brightening observed in the afterglows of high-$z$ GRBs is further sometime enhanced by sporadic spectroscopic observations of the "bumps" which reveal supernova features \citep[e.g. ][]{jin13}. Our Fig. \ref{FigLC} shows a faint optical light curve of ATLAS17aeu suggesting $z>$ 0.3 and the signature of an emerging supernova which starts to outshine the GRB afterglow from $\sim$10-12 days. 

In order to have additional information about the possible SN component we rescaled the absolute $r$-band magnitudes of ATLAS17aeu to the distance of several well-known Type Ib/c SNe, and compared our data with their light curves that cover a wide range of brightness (Fig. \ref{FigDistMod}). This results in a possible range of distances that can explain the observed late time afterglow re-brightening (0.28 $\leq z \leq$ 0.72) of ATLAS17aeu. The best match is obtained with SN\,1998bw (a typical Type Ib/c SN associated with the sub-luminous gamma-ray burst GRB\,980425) assuming a redshift of $z \sim$ 0.6 (see Fig. \ref{FigLC}). We note that even in the case of the match with the brightest known SN associated with a GRB (SN\,2003lw) we obtain $z \sim 0.7$. The hypothesis of a higher redshift would require a much more luminous SN, that has never been observed in association with a long GRB.

   \begin{figure}[!ht]
   \centering
   \includegraphics[width=7.3cm,height=9.1cm,angle=270]{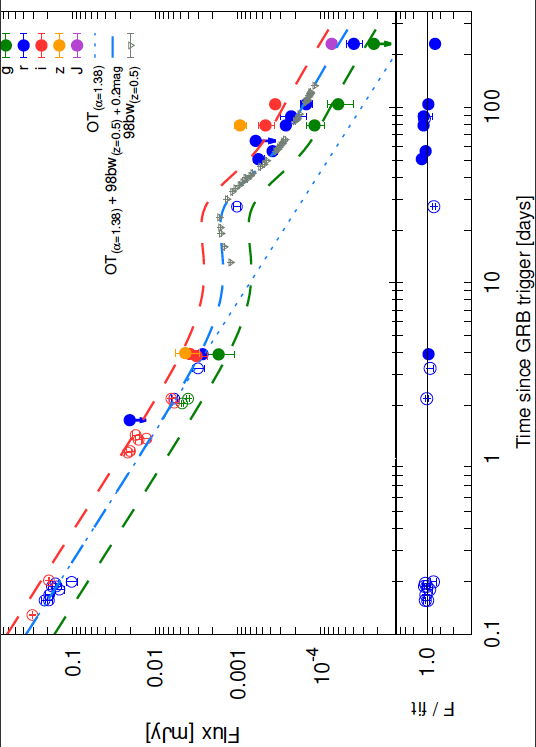}
    \caption{The optical light curve for ATLAS17aeu. Filled points identify our data while empty symbols are data from the literature. The power-law decay of the optical afterglow (blue dashed line) and the SN\,1998bw (gray open triangles) template at $z$=0.5 are shown. The overall fit to the light curve, assuming SN\,1998bw as a template (fainter by 0.2 mag) is shown with blue solid dashed line. The overall fit is then shifted arbitrarily to guide the eye and match the $g$ (green solid dashed line) and $i$ (red solid dashed line) band data.}
    \label{FigLC}%
   \end{figure}

\subsection{Spectral analysis}

To investigate the possible connection between ATLAS17aeu and GRB\,170105A, the early time spectrum obtained with GTC was compared with several Type Ib/c supernova templates. A good match is found with the Type Ic SN\,2003jd \citep{valenti08}, reproducing well the overall shape of the spectrum. A satisfactory comparison is also obtained with SN\,2006aj \citep{pian06,mira06,sol06}, a well-studied supernova (Fig. \ref{FigSpec}) associated with an under-energetic long-duration GRB\,060218 \citep{campana06}. Our analysis showed a possible supernova (SN) signal if a redshift $z \sim 0.6$ is assumed. 

A similar value for the redshift is also found when comparing the late time LBT spectrum of the host galaxy with the template of a star-forming galaxy, by identifying several Balmer transitions at redshift $z \sim 0.623$. The red region of the spectrum (which is the one with the higher signal-to-noise ratio) shows a correspondence between the observed lines (H$\eta$, H$\&$K, H$\delta$ and Gband+H$\gamma$+Fe4383 \AA) and the model. These spectral comparisons indicated a plausible redshift for ATLAS17aeu of $z \sim 0.6 \pm 0.1$.

   \begin{figure}[!ht]
   \centering
    \includegraphics[width=9.0cm,height=8.5cm]{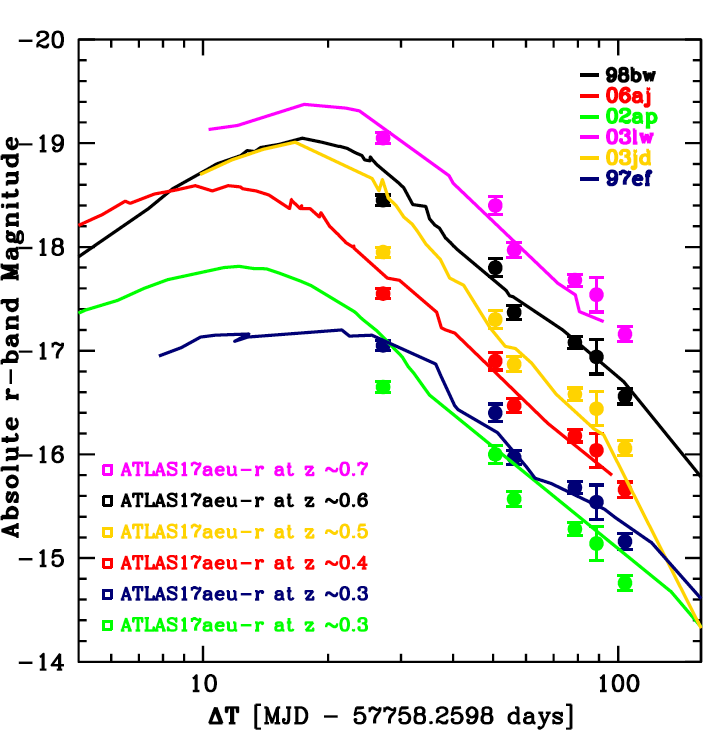}
   \caption{Comparison between the absolute $r$-band magnitudes of several well known Type Ib/c SNe and ATLAS17aeu observations. Times are days since GRB\,170105A trigger time and magnitudes have been $k$-corrected. From the match with each SNe light curve we estimated the possible distance modulus (DM) of ATLAS17aeu and infer the possible range of redshifts (0.3 $\leq z \leq$ 0.7) for the transient.}
    \label{FigDistMod}%
    \end{figure}

   \begin{figure*}[!ht]
%   \centering
    \includegraphics[width=7.5cm,height=9.55cm,angle=90]{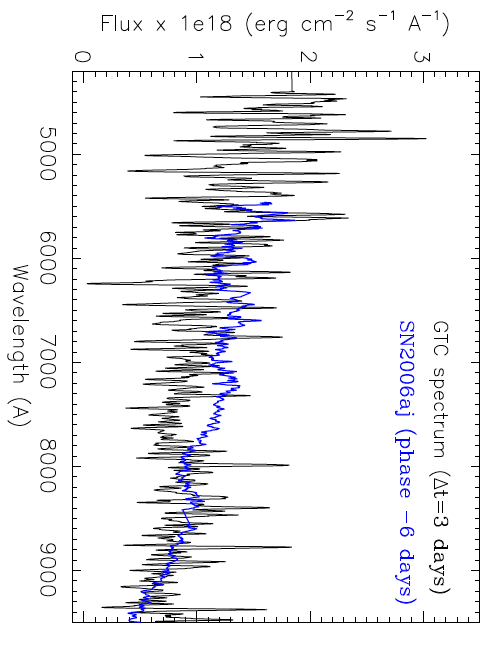}
    \includegraphics[width=7.5cm,height=9.55cm,angle=90]{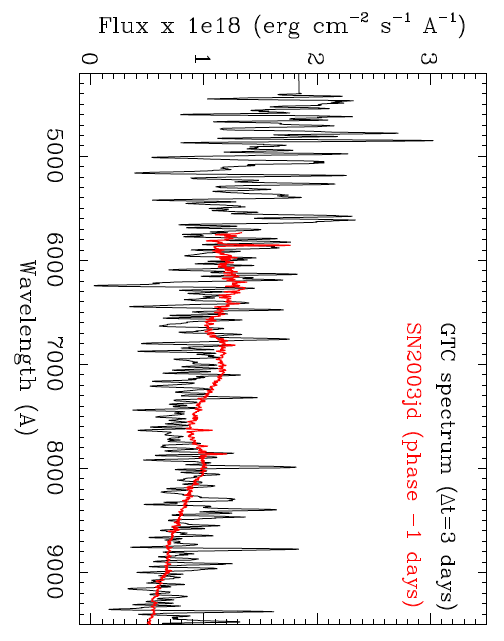}
   \caption{{\it Left}: Comparison between the early time GTC spectrum of ATLAS17aeu and the template (assuming a redshift $\sim 0.6 \pm 0.1$) of the well-studied Type Ib/c SN\,2006aj \citep{pian06}, at a phase of about six days before B-maximum light. {\it Right}: Same assumption as for the left panel, using the template of the Type Ic SN\,2003jd \citep{valenti08}, at a phase of about one day before B-maximum light.}
    \label{FigSpec}%
    \end{figure*}

\subsection{Spectral energy distribution of the afterglow}

In order to study the transient, we modeled and investigated the data-set separately at different wavelength ranges: radio, optical and X-rays. We then interpolate the data to two common epochs and performed a broad band analysis.

We first fitted the radio light curve, for which there are observations at different frequencies, with most of the data at 15.5 GHz and only a couple of detections at 7.4 and 5.0 GHz \citep{bhalerao}. The light curve has a different behaviour in the different bands, and it is decaying more rapidly at higher frequencies (Fig. \ref{Figlcrox}, left panel). When we modeled the data with a power-law fit, we obtained decaying indices of $\alpha_{\rm 15.5 GHz} = 0.66 \pm 0.04$, $\alpha_{\rm 7.4 GHz} = 0.4$ and $\alpha_{\rm 5.0 GHz} = 0.05$ at 15.5, 7.4 and 5.0 GHz, respectively. Note that for the latter two indexes the data points are as many as the parameters, and no uncertainty can be provided. The different decays in the radio bands might be due to colour evolution, which can only be explained by the presence of a spectral break moving from higher to lower frequencies.

Then, to study the behaviour in the optical band, we considered the data up to 5 days after the trigger, when only the afterglow is contributing to the observed flux. Data were corrected for foreground Galactic extinction. The light curve is best followed up in the $r$ and $i$ filters, with $g$ and $z$ bands data starting only 2 days after the trigger. We studied the $griz$ spectral energy distribution of the afterglow at 3.93 days for which we have detections in all the optical bands. We find that the data are best modeled by a power-law with spectral index $\beta_{\rm opt}=1.21 \pm 0.01$ and negligible dust extinction along the line of sight. Afterwards, we modeled all optical light curves together and sampled the time and wavelength plane with a two variable power-law $F(\nu,t) \propto t^{-\alpha_{\rm opt}} \nu^{-\beta_{\rm opt}}$, and fixed $\beta_{\rm opt}$ to the value reported above. This approach is only possible given the negligible dust extinction. In this way, we find an optical decay common to all optical bands of $\alpha_{opt}=1.38\pm 0.02$.

Afterward, we studied the $\it Swift$-XRT data. The data span the interval between $\sim$1 and $\sim$15 days and can be best modeled by a single power-law model with a decay $\alpha_{\rm X} = 0.87 \pm 0.24$. The X-ray spectrum is rather poor and can be fitted using \texttt{Xspec v12.9.0} with a simple power-law with $\beta=0.7\pm0.1$, fixed foreground Galactic absorption ($0.66\times10^{21}$~cm$^{-2}$; \citealt{Willingale2013a}) and negligible host gas absorption. 
%However, this value is in disagreement with the optical spectral slope. 

Finally, we modeled all optical, radio and X-ray spectral energy distribution (SED) at the logarithmic-mean time of the XRT observations, i.e. $\sim$ 3.28 days. We also selected another epoch at 2.14 days, for which we have optical $gri$ detections. We interpolated radio and optical data to the first epoch, and radio and XRT data to the second epoch. In the following we fixed the optical spectral slope to the value $\beta_{\rm opt}=1.21$ found above. We modeled the SED at 3.28 days with a double broken power-law, and we find two spectral breaks: a first spectral break in the radio bands at ($0.7\pm0.1$)$~\times~10^{10}$ Hz and a second break between radio and optical bands at ($1.0\pm0.1$)$~\times~10^{12}$ Hz. Following the standard synchrotron theory under slow cooling regime \citep{Sari1998a}, we identify the first break in the radio to be the absorption frequency $\nu_a$ and the break  between optical and radio to be the injection frequency $\nu_m$. The slope between $\nu_a$ and $\nu_m$ is fixed to the value of $1/3$. %The X-ray emission is of different origin  a cooling break may lie between optical and X-rays. 
It is important to note here that the first break $\nu_a$ is evolving with time and that the decay below the break is almost negligible. This behaviour can be interpreted within the jet scenario \citep{Sari1999a} and the slow cooling regime, which indeed predicts $\alpha=0$ for $\nu<\nu_{a}$ and $\nu_a\propto t^{-1/5}$. Thus, to obtain the model at 2.14 days we followed \citet{Sari1999a} and used the relations $\nu_a\propto t^{-1/5}$ and $\nu_m\propto t^{-2}$.

In  Fig.~\ref{FigSED} we show the radio, optical and X-ray SEDs at different epochs. The fit is acceptable, but we must note that the model does not perfectly match the optical data at the first epoch and the jet scenario would predict more rapid decay in optical and X-rays bands. This suggests the presence of a second break between optical and X-rays and thus a more sophisticated analysis is needed to fully understand the afterglow behaviour. This can be seen in Fig.~\ref{Figlcrox} (right panel) which shows radio, optical, and X-rays light curves. In particular, while the radio and the X-rays light curves agree within 1$\sigma$ (due to the large uncertainty of $\alpha_{\rm X}$), the optical light curve is not consistent with the others. We interpret this as the presence of another spectral break between optical and X-ray bands. 
%In Table ~\ref{tab:TableSED}, we outline the best-fit results
%of the energy breaks and the different spectral slopes. 
An evolving break between optical and X-rays can be seen in the jet scenario without sideways expansion if the circumburst medium has a wind profile, and the synchrotron cooling frequency $\nu_c$ lies in between optical and X-rays bands implying $\beta_{\rm X}-\beta_{opt}=0.5$, consistent with the values reported above \citep[e.g.,][]{Racusin2009,Schulze2011}. It is also expected that $\alpha_{opt}-\alpha_{\rm X}=0.25$, and indeed doing so the two decay indexes are consistent within 2$\sigma$. 
%To test this interpretation we simultaneously modeled the optical to X-rays SED. This could also allow us to constrain the redshift together with optical extinction and dust absorption. Using \texttt{Xspec v12.9.0}, we modeled the SED using a broken- power-law model and considering circumburst dust extinction and gas absorption(\texttt{zdust,ztbabs}) and fixing Galactic absorption like above. The optical to X-ray fit gives a spectral index in the optical bands ($1.02\pm0.03$) smaller than when considering optical data only, and in agreement only within 6 sigmas.  Moreover, we notice that if a break exists, it has to lie very close or within the XRT spectral range. Dust extinction is negligible (as noted above) and X-ray data are too poor to constrain absorption and redshift and the possible break. 
%Alternatively, the X-rays can be dominated by a different mechanism. Indeed, the X-ray decay is typical of a plateau that is difficult to explain within the collapsar scenario, because it requires a long activity of the central engine. 
%The required energy may come from the spin-down activity of a magnetar formed during the collapse \citep[e.g.,][]{ZhangMeszaros2001}.

%----------------------------------- SED fit figures
 \begin{figure}[!ht]
 \centering
  \includegraphics[width=7.0cm,height=9.0cm,angle=270]{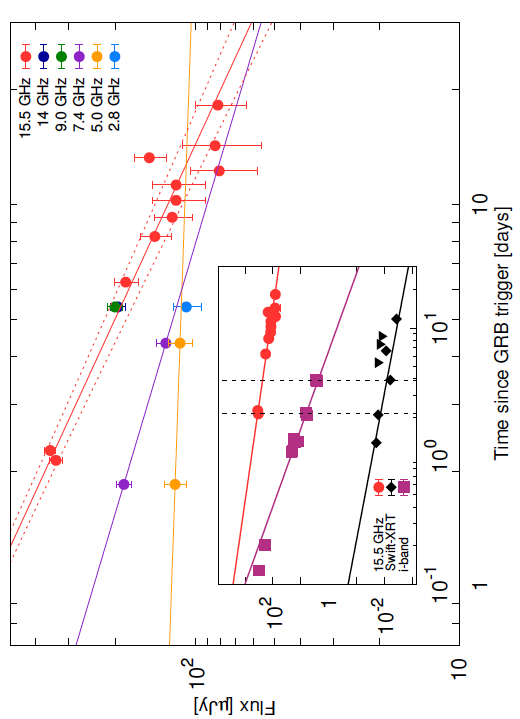}
  \caption{Radio band light curves of ATLAS17aeu at different frequencies. The data at 5.0 (gold), 7.4 (purple) and, 15.5 (red) GHz are fitted  with simple power-laws. {\it Inset plot}: radio (red), optical (plum) and X-rays (black) light curves. Black triangles represent upper limits in the X-rays band. They can all be modeled with simple power laws. Dashed vertical lines represent the selected times for the spectral energy distribution fitting shown in Fig. \ref{FigSED}.}
    \label{Figlcrox}%
    \end{figure}

 \begin{figure}[!ht]
 \centering
  \includegraphics[width=7.0cm,height=9.0cm,angle=270]{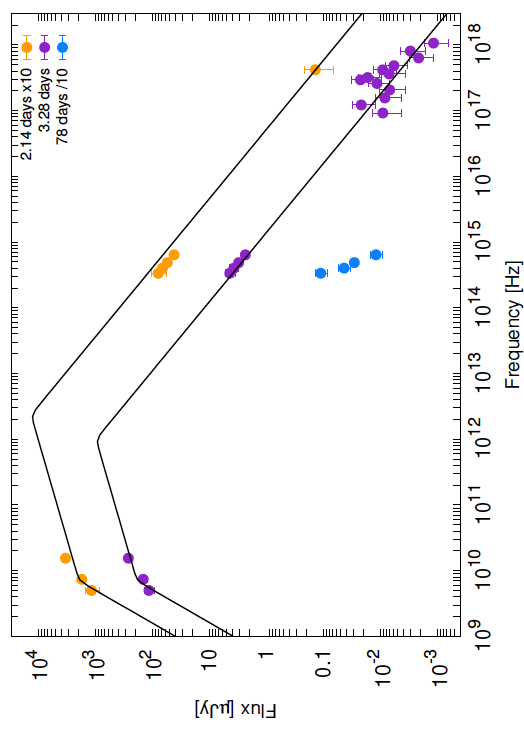}
  \caption{Radio, optical and X-rays data at 3.28 (purple) and 2.14 days (gold) of ATLAS17aeu. The model described in the text is also shown at both epochs. Data at 78 days (cyan) are clearly dominated by the supernova emission and we show them only for comparison with previous epochs.}
    \label{FigSED}%
    \end{figure}

\subsection{Spectral energy distribution of the host galaxy}

%The nearby face on spiral galaxy (marked SDSS J091312.36+610554.2 in Fig.\ref{FigCampo}) has a redshift of $z=0.199$ \citep{stalder17}. If at this redshift the source would be offset from the galaxy by $\sim 78$ kpc, and have an absolute magnitude of $M_{F606W} = -12$. The offset is not unusual for a globular cluster, but the absolute magnitude is approximately a magnitude brighter than any system in the Harris catalog, and is comparably bright to the most luminous globular clusters seen around M87 \citep{waters06}. This may argue against the object as a globular cluster, although it is unusual for GRB hosts to be unresolved in {\em HST} imaging. If binary black holes were produced dynamically in globular clusters, as has been suggested extensively REFS, then they may preferentially be found in the brightest examples where the interaction rates are highest. 

We used the photometric SED-fitting code LePHARE\footnote{http://www.cfht.hawaii.edu/\~arnouts/LEPHARE.} \citep{Arnouts1999,Ilbert2006} to determine host-galaxy parameters from the detections. After fixing the redshift to the most plausible value found in our spectral and temporal analysis ($z \sim$ 0.62, see Sections 4.1 and 4.2), we found that the host is a low-mass (M = 10$^{8.2_{-0.2}^{+0.2}}$\,M$_{\odot}$), galaxy with low global extinction (E$_{\rm B-V} \sim 0.2$~mag using \citealt{Calzetti2000} extinction law), and low star-formation rate (SFR = 0.9$_{-0.4}^{+1.5}$\,M$_{\odot}$~yr$^{-1}$). The inferred low mass is in agreement with the mass of typical long GRB hosts at these redshifts \citep{vergani15}. 

Despite the low SFR, the low mass does not qualify this galaxy as an early type, which would be very unusual for the host of a long GRB \citep[but see][]{Rossi2014a}. In fact, the main stellar population is moderately young (age = $0.3_{-0.2}^{+0.6}$ Gyr) and the galaxy has a high specific SFR of $10^{-8.3_{-0.4}^ {+0.5}}$~yr$^{-1}$ in agreement with other GRB hosts and star-forming galaxies \citep[][]{Hunt2014a,jure16}. The result of our fit is shown in Fig.~\ref{fig:sedhost}. It is worth noting that if we do not fix the redshift we can use photometric data to constraint it between $0.4<z<2.8$, which is well expected given the featureless SED and still inconsistent with the inferred distance of GW\,170104.

To give some indications on the properties of the host galaxy, in Fig. \ref{fig:colours} we also plot the (F390-F140W) colour versus the (F606-F140W) colour of the host galaxy, together with stellar population models. The integrated colour predictions shown in the figure are based on the Stellar Population Tools (SPoT) code for single-age, single-metallicity stellar population (SSP) models \citep{brocato99,raimondo09}, updated for this study using higher total stellar masses, and new spectral libraries for cool and hot stars. Models suggest that the main component of the stellar population in the galaxy is as young as few ten of Myr or younger, in agreement with a high specific SFR \citep[e.g. ][]{feulner05}. This is mildly in agreement with the results of our photometric host-galaxy SED-fitting. The model and data uncertainties do not permit clear indications on the chemical composition of the stellar content.

\begin{figure}[!ht]
\centering
\includegraphics[width=9.3cm,height=7.1cm]{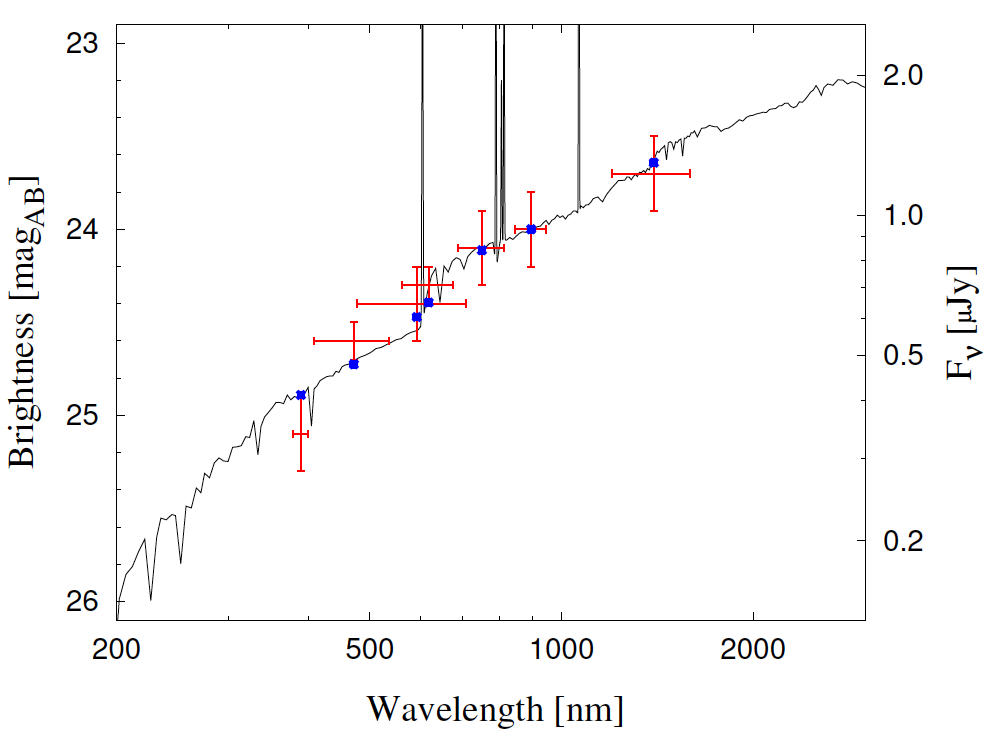}
   \caption{LePHARE fit to the magnitudes of the host galaxy of ATLAS17aeu/GRB\, 170105A with the redshift fixed to be the same as the spectroscopic one. The photometric points are highlighted in red and the blue marks represent the photometry values as determined by the synthetic SED. The fit is acceptable with $\chi/N_{filters} = 2.97/7$. For specific values, see text.}
    \label{fig:sedhost}%
    \end{figure}

\begin{figure}[!ht]
\centering
  \includegraphics[width=8.0cm,height=7.0cm]{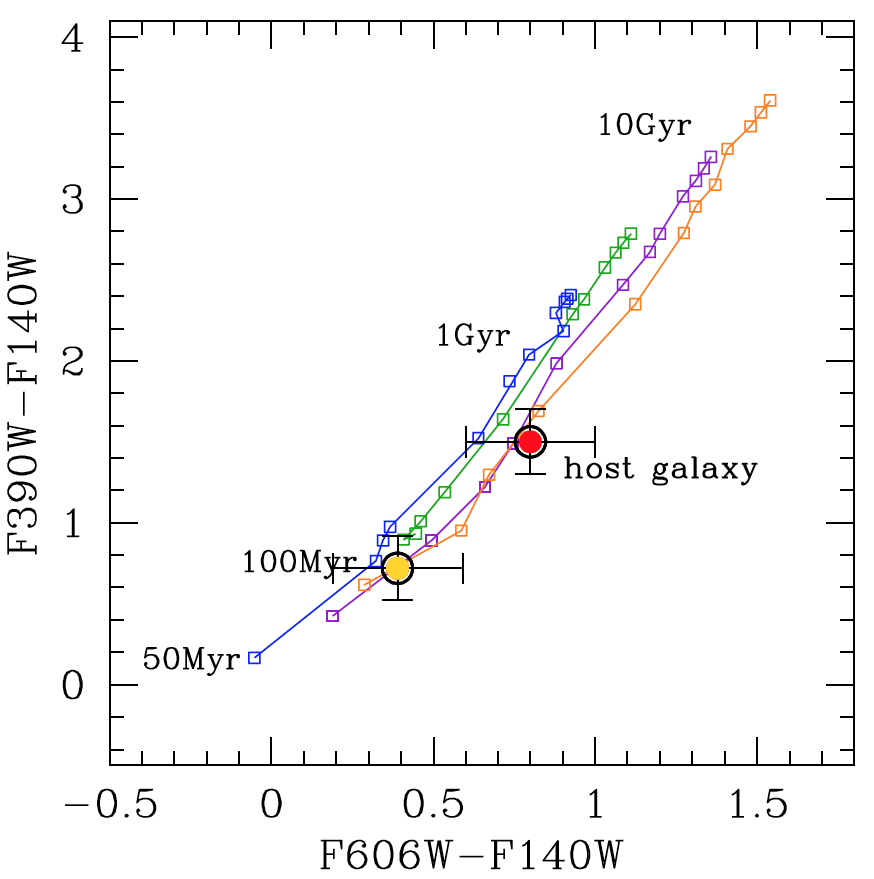}
   \caption{Distance-independent two-colour diagram. Simple stellar population colours from the SPoT code are compared to the measured host-galaxy colors (filled yellow circle). Lines and small squares refer to models with metallicity [Fe/H]=-0.7 (blue), [Fe/H]=-0.4 (green), [Fe/H]=0.0 (violet), [Fe/H]=+0.4 (orange), respectively. Indicative ages are also labeled, from 50 Myr to 14 Gyr. The host galaxy is plotted as a filled red (observed) and yellow (de-reddened) circle. The last value is obtained by applying the value E(B-V)=0.2). No $k$-correction is applied.}
    \label{fig:colours}%
    \end{figure}

%--------------------------------------------------- Photometry table
   \begin{table}[!ht]
      \caption[]{Imaging log for ATLAS17aeu. Different columns correspond to: modified Julian date (1) and $\Delta$t corresponding to the mid time of the observation (2), total exposure time (3), filter identification (4), calibrated AB magnitude not corrected for Galactic extinction (5), and Telescope used for the observation (6).}
         \label{LogTab}
\begin{minipage}{6cm}
\begin{tabular}{cccccl}
\hline \hline
 MJD & $\Delta$t\footnote{This time is estimated from T$_{\rm 0}$.} & t$_{\rm exp}$ & Filt. & Mag (err) & Tel.\\
 $[$d$]$ & $[$d$]$ & $[$min$]$ & & \\
 \hline \hline
 57762.179 & 3.92 & 8 & $g$ & 23.53 (0.08) & GTC \\
  57836.964 & 78.7 & 10 & $g$ & 26.34 (0.26) & GTC\\ 
  57862.194 & 103.9 & 60 & $g$ & 27.05 (0.31) & LBT\\
  57987.330 & 229.1 & 45.6 & {\small F390W} & $>$ 28.1 & HST\\
  \hline
  57820.050 & 61.8 & 80 & $V$ & $>$22.6 & TNG \\
  \hline
 57759.914 & 1.65 & 15 & $r$ & $>$20.7 & Asiago\\
 57762.187 & 3.93 & 10 & $r$ & 23.13 (0.06) & GTC\\
 57808.852 & 50.6 & 70 & $r$ & 24.65 (0.18) & TNG\\
 57814.436 & 56.2 & 65 & $r$ & 25.03 (0.15) & TNG\\
 57822.567 & 64.3 & 90 & $r$ & $>$ 25.0 & TNG\\
 57836.973 & 78.7 & 10 & $r$ & 25.43 (0.12) & GTC\\ 
 57846.960 & 88.7 & 140 & $r$ & 25.48 (0.29) & TNG\\
 57862.194 & 103.9 & 60 & $r$ & 26.05 (0.15) & LBT \\
 57987.330 & 229.1 & 24.4 & {\small F606W} & 27.64 (0.21) & HST\\
 \hline
 57761.093 & 3.84 & 16 & $I$ & 22.52 (0.32) & TNG \\
 57762.197 & 3.94 & 10 & $i$ & 22.85 (0.10) & GTC \\
 57836.953 & 78.7 & 12 & $i$ & 24.98 (0.26) & GTC\\
 57862.178 & 103.9 & 60 & $i$ & 25.13 (0.12) & LBT\\
 \hline
 57762.205 & 3.94 & 7 & $z$ & 22.67 (0.13) & GTC\\ 
 57836.980 & 78.7 & 8 & $z$ & 24.01 (0.24) & GTC\\ 
 \hline
  57987.330 & 229.1 & 16.8 & {\small F140W} & 25.87 (0.14) & HST\\
   %\hline
\end{tabular}
\end{minipage}
\end{table}

%

%--------------------------------------------------- Spectroscopy table
   \begin{table}
      \caption[]{Spectroscopic log for ATLAS17aeu. }
      %Columns are: modified Julian date (1) and $\Delta$t (2) corresponding to the mid time of the observation, total exposure time (3), filter identification (4), calibrated magnitude (5) not corrected for Galactic extinction, and Telescope used for the observation (6).}
         \label{LogTab2}
\begin{minipage}{6cm}
\begin{tabular}{cccccc}
\hline \hline
 MJD & $\Delta$t\footnote{This time is estimated from T$_{\rm 0}$.} & t$_{\rm exp}$ & grism & seeing & Tel.\\
 $[$d$]$ & $[$d$]$ & $[$min$]$ & & [$^{\prime\prime}$] & \\
 \hline
 57761.110 & 2.89 & 2x20 & R1000B & 1.5 & GTC \\
 57761.153 & 2.93 & 2x20 & R2500I & 1.5 & GTC \\
 57858.210 & 99.99 & 3x30 & blue & 0.9 & LBT \\
 57858.210 & 99.99 & 3x30 & red & 0.9 & LBT \\
 57873.903 & 115.68 & 2x20 & R1000B & 0.8 & GTC \\
 57873.924 & 115.70 & 1x20 & R2500I & 0.8 & GTC \\
 58142.431 & 384.21 & 4x10 & blue & 1.2 & LBT \\
 58142.431 & 384.21 & 4x10 & red & 1.2 & LBT \\
 58196.273 & 438.05 & 8x10 & dual-grating & 0.7 & LBT \\
\end{tabular}
\end{minipage}
\end{table}

   \begin{table}[!ht]
\centering
\caption[]{Multi-band photometry of the host galaxy. Columns are: filter identification (1), calibrated  AB magnitude not corrected for Galactic extinction (2), and Telescope used for the observation (3)}
         \label{Loghost}
%\begin{minipage}{6cm}
\begin{tabular}{ccc}
\hline \hline
Filter & Magnitude (error) & Telescope\\
% & & \\
 \hline \hline
%  57987.6 & 255 & 2736 & {\small F390W} & 25.2 (0.2) & HST \\
%  57987.7 & 255 & 1460 & {\small F606W} & 24.5 (0.2) & HST\\
%  57987.7 & 255 & 1006 & {\small F140W} & 23.7 (0.2)  & HST\\
%  57862.1 & 104 & 1800 & $g$ & 24.69 (0.11) & LBT \\
%  57862.1 & 104 & 1800 & $r$ & 24.38 (0.08) & LBT\\
%  57862.1 & 104 & 1800 & $i$ & 24.14 (0.06) & LBT\\
%  XXX & XXX & XXX & $z$ & 24.02 (0.04) & GTC\\
   %\hline
  {\small F390W} & 25.2 $\pm$ 0.2 & HST \\
  $g$ & 24.69 $\pm$ 0.11 & LBT \\ 
  {\small F606W} & 24.5 $\pm$ 0.2 & HST\\
  $r$ & 24.38 $\pm$ 0.08 & LBT\\
  $i$ & 24.14 $\pm$ 0.06 & LBT\\
  $z$ & 24.02 $\pm$ 0.04 & GTC\\
  {\small F140W} & 23.7 $\pm$ 0.2  & HST\\
\end{tabular}
%\end{minipage}
\end{table}

\section{Conclusion}

Our optical observations allowed us to comprehensively describe the temporal behaviour of the unusual transient ATLAS17aeu from early to very late phases. The detection of spectral absorption features reminiscent of broad-lined Ic supernova confirms that ATLAS17aeu is indeed the optical afterglow of the long-duration under-energetic GRB\,170105A, and definitely not associated with the gravitational wave signal GW\,170104, which was due to a binary BH merger \citep{LVC2017a}. 

The presence of the supernova is in fact confirmed at early times in our first spectrum ($\sim 3$ days after the burst event) and at later times by the typical bump in the light curve already seen in many other light curves of GRBs connected SNe. Despite the fact that the redshift for this event is not strongly constrained by the data, we can confidently define a small range of possible values, that is $z \simeq 0.5 \pm 0.2$. The temporal behaviour of such a supernova is similar to the observed evolution of the prototype supernova associated with long GRBs (SN\,1998bw), peaking at similar time after the burst event ($\sim 20$ days). In fact, as for SN\,1998bw, that was associated with a sub-luminous gamma-ray burst (GRB\,980425), also ATLAS17aeu resulted to be associated with a long under-energetic event (GRB\,170105A). All our observations including the host galaxy ones point to the scenario of a long GRB at $z \sim$ 0.5 unrelated to gravitational wave signal.

By assuming the fluence measured by {\it Konus}-WIND and, based on the soft spectrum inferred from the measurements by {\it Konus}-WIND, POLAR and {\it AstroSat}-CZTI, a rest-frame spectral peak energy E$_{\rm p}$ of 50$\pm$25 keV, we find that GRB\,170105A would be consistent with the E$_{\rm p}$-E$_{\rm iso}$ correlation of long GRBs \citep{amati02,amati06} only for $z>$~0.4-0.5 (implying an isotropic energy for this event of E$_{\rm iso} \gtrsim 2~\times$ 10$^{51}$~erg). This finding further supports the above conclusion that this event came from a larger distance with respect to GW\,170104, and is well consistent with its association with a supernova at $z \sim$ 0.5.

\begin{acknowledgements}
We thank P. Shawhan for useful comments. AM acknowledges the support from the ASI grant I/004/11/3. AR acknowledges support from Premiale LBT 2013. AJCT thanks the Spanish Ministry Project AYA2015-71718-R (including FEDER funds). LT, SB are partially supported by the PRIN-INAF 2016 with the project "Toward the SKA and CTA era: discovery, localisation, and physics of transient sources". Partially based on observations collected at Copernico 1.82m telescope (Asiago, Italy) of the INAF - Osservatorio Astronomico di Padova. JH was supported by a VILLUM FONDEN Investigator grant (project number 16599). AFV is thankful to the Russian Science Foundation (grant 14-50-00043). This work made use of observations obtained with the Italian 3.6m Telescopio Nazionale Galileo (TNG) and the 10.4m Gran Telescopio Canarias (GTC), operated on the island of La Palma by the Fundaci\'{o}n Galileo Galilei of the Instituto Nazionale di Astrofisica (INAF) at the Spanish Observatorio del Roque de los Muchachos of the Instituto de Astrof\'{i}sica de Canarias, and also of observations made with the 8.4m Large Binocular Telescope (LBT). The LBT is an international collaboration among institutions in Italy, United States, and Germany. LBT Corporation partners are: Istituto Nazionale di Astrofisica, Italy; The University of Arizona on behalf of the Arizona university system; LBT Beteiligungsgesellschaft, Germany, representing the Max-Planck Society, the Astrophysical Institute Potsdam, and Heidelberg University; The Ohio State University; and The Research Corporation, on behalf of The University of Notre Dame, University of Minnesota, and University of Virginia. We thank the TNG staff, in particular G. Andreuzzi, G. Mainella, A. Harutyunyan, and the LBT staff, in particular A. Gargiulo, for their valuable support with TNG and LBT observations and data reduction. We also acknowledge INAF financial support of the project ”Gravitational Wave Astronomy with the first detections of adLIGO and adVIRGO experiments.

\end{acknowledgements}


\end{document}